%% file: paper.tex
\documentclass[acmsmall, screen]{acmart}
\usepackage{amsmath,amsfonts}
\usepackage{algorithmic}
\usepackage{graphicx}
\usepackage{textcomp}
\usepackage{xcolor}
\usepackage{color}
\usepackage{listings}
\usepackage{subcaption}
\usepackage{booktabs}
\usepackage{tcolorbox}

\usepackage{multirow}

\AtBeginEnvironment{minted}{\singlespacing%
    \fontsize{6}{6}\selectfont}
\usepackage{relsize}
\usepackage[inline]{enumitem}

\usepackage{tikz}

\AtBeginDocument{%
  \providecommand\BibTeX{{%
    \normalfont B\kern-0.5em{\scshape i\kern-0.25em b}\kern-0.8em\TeX}}}

\newcommand{\B}{B}
\newcommand{\CB}{CB}
\newcommand{\ea}{\textit{et al.}}

\newcommand{\smallsection}[1]{\noindent {\bf \underline{#1}}.\hspace{1mm}}

\usepackage{soul}

\newcommand{\rqone}{How accurate are CodeLMs at recommending code updates?}
\newcommand{\rqtwo}{ What is the syntactical correctness of the code generated by CodeLMs?}
\newcommand{\rqthree}{Why do CodeLMs perform differently in time-wise and time-ignore scenarios}
\newcommand{\rqfour}{How do the method size and update size impact the accuracy of CodeLMs?}
\newcommand{\rqfive}{What types of code updates are learned and predicted by CodeLMs?}

\definecolor{dkgreen}{rgb}{0,0.6,0}
\definecolor{gray}{rgb}{0.5,0.5,0.5}
\definecolor{mauve}{rgb}{0.58,0,0.82}

\newcommand{\lstbg}[3][0pt]{{\fboxsep#1\colorbox{#2}{\strut #3}}}
\definecolor{codegreen}{rgb}{0,0.6,0}
\lstdefinelanguage{diff}{
	frame=single,
	breaklines=true, 
	basicstyle=\ttfamily\small,
	morecomment=[f][\color{red}]{---}, 
	morecomment=[f][\color{codegreen}]{+++},
	morecomment=[f][\lstbg{red!20}]{-\ },
	morecomment=[f][\lstbg{green!20}]{+\ },
	morecomment=[f][\color{blue}]{@@},
}

\newcommand{\find}[1]{
\begin{tcolorbox}[leftrule=1mm,toprule=0mm,bottomrule=0mm,left=1pt,right=2pt,top=2pt,bottom=2pt]
\em #1
\end{tcolorbox}
}

\begin{document}

\title{Automatically Recommend Code Updates: Are We There Yet?}

\author{Yue Liu}
\email{yue.liu1@monash.edu}
\author{Chakkrit Tantithamthavorn}
\email{chakkrit@monash.edu}
\author{Yonghui Liu}
\email{Yonghui.Liu@monash.edu}
\affiliation{%
  \institution{Monash University}
  \streetaddress{Wellington Road}
  \city{Clayton}
  \state{Victoria}
  \country{Australia}
}

\author{Patanamon Thongtanunam}
\email{patanamon.t@unimelb.edu.au}
\affiliation{%
  \institution{The University of Melbourne}
  \streetaddress{Grattan Street}
  \city{Parkville}
  \state{Victoria}
  \country{Australia}
}

\author{Li Li}
\email{lilicoding@ieee.org}
\affiliation{%
  \institution{Beihang University}
  \city{Beijing}
  \country{China}
}

\begin{abstract}
In recent years, large pre-trained Language Models of Code (CodeLMs) have shown promising results on various software engineering tasks. 
One such task is automatic code update recommendation, which transforms outdated code snippets into their approved and revised counterparts.
Although many CodeLM-based approaches have been proposed, claiming high accuracy, their effectiveness and reliability on real-world code update tasks remain questionable.
In this paper, we present the first extensive evaluation of state-of-the-art CodeLMs for automatically recommending code updates.
We assess their performance on two diverse datasets of paired updated methods, considering factors such as temporal evolution, project specificity, method size, and update complexity.
Our results reveal that while CodeLMs perform well in settings that ignore temporal information, they struggle in more realistic time-wise scenarios and generalize poorly to new projects. Furthermore, CodeLM performance decreases significantly for larger methods and more complex updates. Furthermore, we observe that many CodeLM-generated "updates" are actually null, especially in time-wise settings, and meaningful edits remain challenging. 
Our findings highlight the significant gap between the perceived and actual effectiveness of CodeLMs for real-world code update recommendation and emphasize the need for more research on improving their practicality, robustness, and generalizability.

\end{abstract}



\maketitle

\input{sections/1_introduction}

\input{sections/2_background}
\input{sections/3_approach}
\input{sections/4_experiment_setup}

\input{sections/5_result}

\input{sections/discussion}
\input{sections/threats}
\input{sections/6_relatedwork}
\input{sections/7_conclusion}

\bibliographystyle{IEEEtranS}
\bibliography{reference}


\end{document}

%% file: sections/1_introduction.tex
\section{Introduction}
In recent years, large pre-trained language models for code generation and understanding have gained significant popularity~\cite{hou2023large}. 
These models, known as Language Models of Code (CodeLMs), have demonstrated promising results across a wide range of software engineering tasks~\cite{hou2023large, fan2023large}.
A number of CodeLMs such as CodeBERT~\cite{feng2020codebert}, CodeT5~\cite{wang2021codet5}, and CodeGPT~\cite{lu1codexglue} have been released for software engineering tasks, including code search~\cite{gu2018deep}, code generation~\cite{kim2021code}, defect prediction \& localization~\cite{dam2019lessons, huo2019deep}, and code review~\cite{tufan2021towards, tufano2019learning, thongtanunam2022autotransform}. 
While recent efforts have aimed at enhancing the performance of these models, previous studies~\cite{liu2023reliability, liu2023towards, shi2022we} have indicated the lack of standardized and realistic benchmarks that can measure their generalization and robustness. Prior works~\cite{liu2023reliability, liu2023towards, shi2022we} have shown that even popular datasets like CodeSearchNet~\cite{husain2019codesearchnet} and CodeXGlue~\cite{lu1codexglue} suffer from various issues that lead to unreliable performance measures.
For instance, Liu~\ea~\cite{liu2023reliability} discovered severe data duplication problems in several program generation datasets from CodeXGlue~\cite{lu1codexglue}, which resulted in over-optimistic performance.
Moreover, most datasets lack sufficient metadata, such as the source and temporal information, leaving us in the dark about the reasons behind the success or failure of CodeLMs~\cite{she2023pitfalls}. 
As a result, the effectiveness of these models in real-world software development and maintenance scenarios remains an open question.

To better understand and examine the real-world applicability of CodeLMs, we focus on a prevalent task in software maintenance: code updates.
Code updates represent a constant and evolving challenge in software development, as developers must continually integrate dependencies, adapt to new APIs, migrate between frameworks, fix newly discovered bugs, and add features to meet user needs.
Given the frequent and repetitive nature of code updates, CodeLMs could be a valuable tool for automating such tasks, thereby reducing the manual effort required from developers.
However, as highlighted by previous studies~\cite{liu2023reliability, liu2023towards, shi2022we}, the reliability and robustness of CodeLMs on practical software engineering tasks remain questionable.
The question, "\textbf{Can CodeLMs effectively recommend code updates?}" and the factors contributing to their success or failure in recommending code updates, remain unanswered. Therefore, in this paper, we aim to investigate the potential of CodeLMs for automating code updates.
We identify the key challenges and opportunities in this direction with the goal of providing insights that can guide future research towards making automated code updates a reality.

\emph{In this work}, we present the first extensive evaluation of recent CodeLMs in automatically recommending code updates.
To ensure a comprehensive and rigorous assessment, we utilize two unique datasets of paired updated methods. 
The first dataset is extracted from 3,195 Android apps published on Google Play and hosted on GitHub, spanning the period from 2008 to 2022.
The second dataset is an existing benchmark dataset from Pornprasit~\ea~\cite{pornprasit2023d}, which includes code updates from three diverse software projects: Android, Google, and Ovirt.
These datasets provide a rich and diverse source of information, enabling us to rigorously assess the capabilities of CodeLMs in handling real-world code updates.
To determine the effectiveness of CodeLMs in recommending code updates, we train and evaluate various CodeLMs (i.e., CodeT5~\cite{wang2021codet5}, CodeT5+\cite{wang2023codet5+}, CodeBERT\cite{feng2020codebert}, CodeGPT~\cite{lu1codexglue}, UniXcoder~\cite{guo2022unixcoder}) and two baselines (i.e., T5~\cite{raffel2020exploring}, AutoTransform~\cite{thongtanunam2022autotransform}).
Our evaluation methodology is carefully designed to consider various factors that may impact the performance of CodeLMs. 
We take into account the temporal nature of code updates by evaluating the models in both time-wise and time-ignore scenarios. This allows us to assess the models’ ability to capture the evolutionary aspects of code changes. Furthermore, we analyze the performance of CodeLMs across different code update types, method sizes, and update complexities. 
we aim to uncover the strengths (\textbf{good}), limitations (\textbf{bad}), and areas for improvement (\textbf{missing}) of CodeLMs in handling diverse code update scenarios.
We address the following five research questions:

\begin{enumerate}[label={\bf(RQ\arabic*)}]

\item {\bf \rqone}\\
\smallsection{Results} 
CodeLMs perform well in time-ignore scenarios, achieving up to 36.9\% accuracy, and increasing the beam search size can improve their performance. However, they perform poorly in realistic time-wise evaluations and cross-project settings, with accuracy decreases ranging from 58.7\% to 100\%. 

\item {\bf \rqtwo}\\
\smallsection{Results}
State-of-the-art CodeLMs achieve high syntactical correctness rates, often above 90\%, in most datasets and scenarios. However, even with high syntactical correctness rates, CodeLMs still generate syntactically incorrect code in some instances, such as mismatching parentheses. 

\item {\bf \rqthree}\\
\smallsection{Results}
Time-ignore data has substantial overlap (17-52\% of test samples seen in training), while time-wise has much less (2-28\%).
CodeLMs do well on seen samples but struggle with novel and complex ones, especially in time-wise settings. 

\item {\bf \rqfour}\\
\smallsection{Results}
CodeLMs perform better on smaller methods and updates with fewer changed tokens. However, their accuracy decreases significantly as method size and update complexity increase, dropping to 0\% for larger methods and more complex updates.

\item {\bf \rqfive}\\
\smallsection{Results}
Many predicted "updates" are actually null (no changes made), especially in time-wise settings (up to 92\%).
When changes are made, they are often simple, frequent updates. Meaningful edits remain challenging.
\end{enumerate}

In summary, while CodeLMs show potential for automating code updates, our results highlight significant challenges in adapting to new projects, generating valid code, and recommending meaningful changes, particularly in realistic time-wise settings.
We hope our findings will motivate future work on improving the practicality of CodeLM-based code update tools.
The key contributions of this paper are:

\begin{itemize}
    \item Conduct an extensive evaluation of state-of-the-art CodeLMs for the task of automatically recommending code updates;
    \item An in-depth comparative analysis of CodeLM performance across different scenarios and granularities;
    \item Insights into the current challenges facing CodeLMs in this domain, with recommendations for future research directions.
\end{itemize}

\noindent\textbf{\underline{Open Science.}} To support the open science initiative, we publish our dataset and a replication package, which is publicly available in GitHub.\footnote{https://github.com/yueyueL/CodeLM-CodeUpdateEval}

\noindent\textbf{\underline{Paper Organization.}}
Section~\ref{sec:background} presents the background and motivation. 
Section \ref{sec:approach} describes the study design of this empirical study.
Section~\ref{sec:experiment} presents our studied datasets and the experimental setup, while Section \ref{sec:results} presents our research questions and the experimental results. 
Section~\ref{sec:discussion} provides a critical discussion of our findings and their limitations.
Section~\ref{sec:threats} discloses the threats to validity. 
Section~\ref{sec:relatedwork} describes the related work.
Section~\ref{sec:conclusion} concludes the paper.

%% file: sections/2_background.tex
\section{Background}
\label{sec:background}
This section provides an introduction to the concept of automated code recommendation and outlines the motivation for this study.

\subsection{Language Models}
Language models (LMs) play a crucial role in natural language processing (NLP) and artificial intelligence (AI)~\cite{liu2023pre}.
By modeling the probability distribution of sequences of words on a training corpus, LMs can be effectively applied to tasks such as text generation~\cite{kenton2019bert} and understanding~\cite{vaswani2017attention}.
LMs generally consist of two primary components: an encoder that processes the input sequence and generates a numerical, intermediate representation, and a decoder that utilizes this intermediate representation to produce the target sequence one token at a time.
The development of language models has witnessed several significant milestones, from early n-gram models to more sophisticated techniques.
Recurrent neural networks (RNNs) have gained prominence as an influential approach capable of capturing longer dependencies and contextual information~\cite{lecun2015deep}.
However, the Transformer architecture revolutionized the field of language modeling with its use of self-attention mechanisms for handling long-range dependencies and parallelism more effectively~\cite{vaswani2017attention, raffel2020exploring}, opening new doors for AI research.

The introduction of the Transformer architecture sparked a proliferation of large-scale, pre-trained LMs, such as BERT~\cite{kenton2019bert}, T5~\cite{raffel2020exploring}, and GPT-3~\cite{brown2020language}, which have demonstrated remarkable improvements across various NLP tasks.
These models benefit from the utilization of extensive training data and transfer learning techniques. 
By pre-training on large generic datasets through self-supervised tasks, these models acquire a comprehensive understanding of language, which can be further fine-tuned on smaller, specialized datasets for specific tasks.
For example, BERT employs a bidirectional transformer and is pre-trained on 3.3 billion words, yielding outstanding results in tasks such as sentiment analysis and question answering~\cite{minaee2021deep}.

Recently, there has been a growing interest in domain-specific pre-trained models, such as CodeT5~\cite{wang2021codet5} and CodeBERT~\cite{feng2020codebert}, which are specifically designed for programming language (PL) processing tasks.
These models are pre-trained LMs of Code (CodeLMs) on extensive programming code corpora and, akin to their general-purpose counterparts, can be fine-tuned on specialized code-related tasks.
These pre-trained models can be categorized into three groups: encoder-only models, decoder-only models, and encoder-decoder models. 
\textbf{Encoder-based models}, like CodeBERT~\cite{feng2020codebert}, exclusively use a bidirectional transformer encoder with an attention mechanism to learn vectorized embeddings of input code sequences. 
These models are best suited for non-generation downstream tasks, such as code representation and code clone detection~\cite{feng2020codebert}. 
\textbf{Decoder-based models}, including CodeGPT~\cite{lu1codexglue} and Codex~\cite{chen2021evaluating}, employ an autoregressive transformer decoder to generate code sequences. Unlike encoder-only models that focus on input code embeddings, decoder-only models excel in open-ended code generation tasks~\cite{chen2021evaluating}.
Finally, \textbf{encoder-decoder-based models} like CodeT5~\cite{wang2021codet5} feature both a bidirectional transformer encoder and an autoregressive transformer decoder. 
The encoder computes input code embeddings, while the decoder generates code. This flexibility allows encoder-decoder models to accommodate both non-generation and generation downstream tasks~\cite{shi2022evaluation, niu2023empirical, wang2022bridging}.
These code-focused pre-trained models have demonstrated exceptional performance in tasks like code summarization~\cite{shi2022evaluation}, code search~\cite{wang2022bridging}, and code completion~\cite{izadi2022codefill}. 
The development of advanced pre-trained CodeLMs, encompassing both general-purpose and domain-specific varieties, has significantly expanded the horizons of NLP and AI research, extending its applicability to the realm of software engineering.

\subsection{Automated Code Update}
Code updates are an important quality assurance practice in the software development process, aimed at ensuring that newly developed code meets the required standards before integration into the main software repository~\cite{zhao2022towards, tufano2019learning, thongtanunam2022autotransform}.
The software update process involves revising the source code to rectify defects and bugs~\cite{zhao2022towards}, enhance code quality and readability~\cite{li2022automating}, and support new functions~\cite{xia2020android}.
However, during the software development lifecycle, developers are required to frequently update the source code, which is time-consuming and human-intensive.
Given the large number of reviews (e.g., 3,000 reviews per month at Microsoft Bing~\cite{rigby2013convergent}), prior studies have found that it is challenging for code authors to revise their code without introducing new defects while switching contexts and keeping track of other reviews~\cite{czerwonka2015code, kononenko2016code}.
Automated code update recommendation could potentially save developers' effort by automatically applying common revisions from past code reviews to newly-written code, allowing code authors to focus on revising more complex code.
This approach would not only reduce the manual effort required from developers but also help maintain code quality and consistency throughout the development process.
By leveraging the knowledge gained from previous code updates, automated code recommendation systems can provide valuable suggestions and assist developers in efficiently updating their code to meet the required standards.

Recent work has explored the use of language models (LMs) to support automated code update recommendation. 
For example, Tufano~\ea~\cite{tufano2019learning} proposed an RNN Encoder-Decoder architecture to learn how to automatically apply Java code changes implemented by developers during pull requests. 
Their approach involves three main steps: code abstraction, building an LM model, and generating predictions. 
The model learns the relationship between the "before" and "after" code sequences and generates updated code based on this learned mapping. 
AutoTransform\cite{thongtanunam2022autotransform} further improved upon this approach by using a Transformer Encoder-Decoder-based architecture and Byte-Pair Encoding (BPE) subword tokenization to better handle new tokens and long code sequences. 
These LM-based approaches have shown promising results in automating code updates and have the potential to significantly reduce the manual effort required from developers.

\subsection{Motivation}
Despite the advancements in automated code update recommendation using LMs, existing approaches still exhibit certain limitations that warrant further exploration.
For instance, Tufano~\ea’s approach~\cite{tufano2019learning} struggles with handling new tokens in the updated code version and efficiently processing long code sequences.
Although AutoTransform~\cite{thongtanunam2022autotransform} improves upon Tufano~\ea’s work by utilizing a Transformer-based architecture and BPE subword tokenization, its overall accuracy remains relatively low, with only 20.4\% accuracy when using a beam search size of 10.
In real-world scenarios, it is impractical and time-consuming for developers to select the correct recommendation from 10 candidates, making this level of accuracy insufficient for practical use.
Secondly, previous studies have not thoroughly assessed the robustness of their proposed approaches concerning real-world constraints. For example, temporal factors and the relation between training and testing splits over time have not been adequately addressed in real-world settings. Liu~\ea~\cite{liu2022explainable} demonstrate that ignoring time information when splitting train and test data can lead to over-optimistic performance of malware detection classifiers. Additionally, in real-world scenarios, cross-project and within-project prediction settings should be considered~\cite{jing2016improved, she2023pitfalls}. However, these settings have not been discussed in previous studies~\cite{tufano2019learning, thongtanunam2022autotransform}.
Therefore, comprehensive studies are required to understand the capabilities and limitations of advanced state-of-the-art language models for code updates in real-world settings.

Moreover, the reliability and trustworthiness of performance measures of CodeLMs remain questionable~\cite{she2023pitfalls}, making it challenging to discern why CodeLMs fail or succeed. For instance, dataset construction often lacks sufficient transparency and metadata to reliably attribute the failure or success of code update recommendations to specific factors~\cite{shi2022we, she2023pitfalls}. Liu~\ea~\cite{liu2023reliability} demonstrated that the dataset used in the Tufano~\ea~and AutoTransform studies is not reliable, providing overestimated performance measures. Specifically, data duplications exist between the training and testing sets, and overlapping input samples in testing sets inflate performance claims. Without transparent and unbiased evaluations, the real-world effectiveness of CodeLMs on practical tasks like code updates remains uncertain.

To address these limitations and investigate the potential of CodeLMs for automated code updates, we conduct an extensive empirical study examining state-of-the-art CodeLMs on real-world settings.
By curating datasets from the real world and evaluating various CodeLMs under different scenarios, considering factors like update types, temporal robustness, project specificity, and method lengths, we aim to provide the first systematic assessment of applying modern CodeLMs to code update recommendations.
Our work aims to uncover the capabilities and limitations of existing models on this practical task, identifying the good, bad, and missing aspects of their performance.
Furthermore, our study aims to provide insights into the factors contributing to the success or failure of CodeLMs in recommending code updates.

%% file: sections/3_approach.tex
\section{Study Design}
\label{sec:approach}

\subsection{Data Preparation}
To effectively evaluate the performance of CodeLMs in recommending code updates, we have curated two datasets: \textit{AndroZooUpdate} and one existing benchmark dataset from Pornprasit~\ea~\cite{pornprasit2023d}.
Both these datasets include commit dates of their corresponding patches, allowing for a time-wise evaluation.
The combination of these two datasets allows us to comprehensively assess the performance of CodeLMs across diverse software projects and update types.
Table~\ref{tab:section2_data_summary} presents the overview of the studied datasets.

\subsubsection{Pornprasit~\ea~\cite{pornprasit2023d}}
Building upon Tufano~\ea~\cite{tufano2019learning}, which extracted method-level code change transformations mined from code review repositories Gerrit~\cite{gerrit2022} of software projects such as \textit{Google}, \textit{Ovirt}, and \textit{Android}, Pornprasit~\ea~\cite{pornprasit2023d} further retrieved the code update patches with temporal information.
This collection forms a code-to-code corpus, where the source code prior to the pull request is transformed into the target code, reflecting the changes post-review.
This code update transformation dataset has been widely used by prior work~\cite{zeng2022extensive, tufano2019learning, liu2023reliability, pornprasit2023d}.

\input{tables/dataset}

\subsubsection{AndroZooUpdate}
In addition to the existing benchmark dataset, we have also collected our own dataset, \textit{AndroZooUpdate}, to understand the performance of CodeLMs in real-world scenarios. This dataset focuses specifically on code updates in Android applications, which are among the most popular mobile software systems worldwide~\cite{liu2021deep}. By curating this dataset, we aim to assess the effectiveness of CodeLMs in a practical context, providing a valuable resource for researchers and practitioners working on automated code update recommendation systems. We have undertaken the following steps to collect and prepare the dataset for code update recommendations:

To collect code updates, we start with a collection of Android apps from the AndroZooOpen dataset~\cite{liu2020androzooopen}, which comprises 46,521 open-source Android apps released between 2008 and 2020. 
AndroZooOpen provides metadata for app repositories, as well as links to the Google Play pages if the apps are published on Google Play.
In this study, we focus exclusively on open-source Android apps published on Google Play to exclude toys, experimental repositories, or low-quality apps.
As a result, we procured 3,195 open-source Android apps from AndroZooOpen. 
Subsequently, utilizing the provided metadata, we cloned the Git repositories of these apps and identified 1.3 million human-written commits for further extraction.

\begin{figure*}[t]
    \centering
    \includegraphics[width=0.5\textwidth]{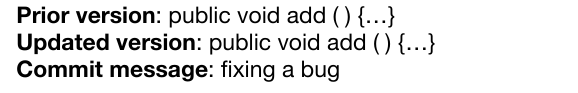}
    \caption{An example of collected data triplet.}
    \label{fig:code_example}
\end{figure*}

To extract all code modifications in the 1.3M commits that we collected, we utilize the \emph{git diff} command in Git to identify which parts of code are updated in the two versions (i.e., before and after the code update). 
In this study, we identify code updates at the method level for two reasons: (1) a method implements a single functionality, providing sufficient context, and (2) file-level code updates may encompass multiple method-level code transformations, with files containing large amounts of text, including significant unchanged code sections that could hinder CodeLM model performance~\cite{tufano2019learning}.
To more accurately evaluate code updates, we removed developers' comments within code blocks. 
We then constructed a list of triplets (i.e., the method block before updates (\emph{prior}), the method block after updates (\emph{updated}), and the corresponding commit messages) to represent the updated methods (i.e., as shown in Figure~\ref{fig:code_example}). 
Ultimately, we extracted 209,346 pairs of updated methods and their corresponding commit messages (i.e., \textit{AndroZooUpdate-L}).
The dataset is represented as a list of triplets, i.e., \emph{$ \{(m_{1},m^{'}{1}, c{1}),...,(m_{N},m^{'}{N}, c{N})\}$}, where \emph{$m_{i}$} denotes the version of a method \textit{prior} to the code update, \emph{$m_{i}^{'}$} signifies the \textit{updated} version, and \emph{$c_{i}$} indicates the corresponding commit message.

\subsection{Pre-trained LMs for Code Recommendation}
Pre-trained CodeLMs are exposed to an extensive variety of programming languages and code samples covering numerous software engineering concepts and constructs during the pre-training phase. 
This enables the model to learn a general understanding of code syntax, semantics, and common programming patterns. 
To gain deeper insight into the impacts of pre-trained CodeLMs on automated code update recommendations, we examined four popular pre-trained CodeLMs for code generation and understanding, namely, CodeBERT~\cite{feng2020codebert} (an encoder-based model), CodeGPT~\cite{lu1codexglue} (a decoder-based model), CodeT5~\cite{wang2021codet5} (an encoder-decoder-based model), CodeT5+~\cite{wang2023codet5+},  and UniXcoder~\cite{guo2022unixcoder} (a unified model).
Additionally, we include T5~\cite{raffel2020exploring}, a model without pre-training in programming languages but designed for natural language processing, to assess the importance of domain-specific pre-training in the context of automated code recommendation.
Tufano~\ea~\cite{tufano2022using} proposed a T5-based pre-trained code model for automated code update activities (called TufanoT5).

\textbf{CodeBERT.} 
CodeBERT is one of the first attempts to pre-train a Transformer-based encoder model with the primary objective of enhancing source code representation learning and comprehension.
Based on the popular BERT architecture~\cite{kenton2019bert}, CodeBERT is an encoder-only model that has been pre-trained on the CodeSearchNet dataset (an open-sourced code corpus containing 4291 projects collected from GitHub)~\cite{husain2019codesearchnet}.
This dataset consists of 2.1 million bimodal NL-PL (comment-function) pairs and 6.4 million unimodal functions without comments, spanning across six programming languages. 

\textbf{CodeGPT.}
CodeGPT~\cite{lu1codexglue} is an innovative pre-trained neural language model that focuses on code generation tasks.
Built upon the GPT architecture, CodeGPT leverages the power of the autoregressive Transformer-based decoder to effectively generate code sequences.
The model undergoes pre-training on Python and Java corpora derived from the CodeSearchNet dataset, encompassing 1.1 million Python functions and 1.6 million Java methods. 
This extensive pre-training process empowers CodeGPT to excel in various code generation tasks, demonstrating its potential in programming language processing and software engineering.

\textbf{CodeT5.}
Built upon the T5 architecture~\cite{raffel2020exploring}, CodeT5 employs an encoder-decoder framework, utilizing bidirectional Transformer encoders and autoregressive Transformer decoders. 
This combination allows the model to not only calculate the embedding of input code but also generate code sequences effectively.
The pre-training dataset for CodeT5 is derived from the CodeSearchNet corpus, as well as C and CSharp programs collected via BigQuery~\cite{BigQuery}, which comprises approximately 5 million natural language instances and 8 million programming language instances.

\textbf{CodeT5+.}
CodeT5+~\cite{wang2023codet5+} is an advanced version of the CodeT5 model, designed to further enhance the capabilities of automated code generation and understanding. 
It extends the CodeT5 model by introducing a mixture of pretraining objectives that cover span denoising, contrastive learning, text-code matching, and causal language modeling, on both unimodal and bimodal multilingual code corpora.
Moreover, it leverages frozen off-the-shelf LLMs to initialize its encoder and decoder modules without training from scratch, and explores instruction-tuning to align the model with natural language instructions.

\textbf{UniXcoder.}
UniXcoder~\cite{guo2022unixcoder} is a unified, cross-modal pre-trained model for programming languages, designed to support both code-related understanding and generation tasks. 
This model effectively combines the strengths of encoder and decoder architectures, enabling it to efficiently learn code representation while performing code generation tasks.
To control the model's behavior, UniXcoder employs mask attention matrices with prefix adapters and harnesses cross-modal content, such as Abstract Syntax Trees (ASTs) and code comments, to enhance code representation. 
The model is pre-trained using a combination of several tasks: masked language modeling (MLM), unidirectional language modeling, denoising autoencoder, and two contrastive learning-related tasks.
This comprehensive pre-training approach allows UniXcoder to excel in a variety of programming language tasks, making it a versatile and powerful tool for code-related understanding and generation applications.

\subsection{Fine-turning CodeLMs}
Fine-tuning is a critical step in adapting pre-trained neural language models to specific tasks or domains, such as automated code recommendation. 
After pre-training on large-scale, diverse datasets, CodeLMs possess a generalized understanding of language structure and patterns. 
However, to excel in targeted tasks, these models often require additional training on smaller, specialized datasets relevant to the task.
In our study, we fine-tune pre-trained CodeLMs to learn automated code recommendations by transforming initial methods into updated versions. 
We use $X_{i} = [x_{1}, ..., x_{n}]$ to denote the initial version and $Y_{i} = [y_{1}, ..., y_{m}]$ to represent the updated version. 
In line with previous studies~\cite{thongtanunam2022autotransform, liu2020multi}, we fine-tune the models on the related training dataset by minimizing the cross-entropy losses:
\begin{equation*}
\mathcal{L}(\theta) = \sum_{i=1}^{N} \log P(Y_{i} | X_{i}; \theta),
\end{equation*}
where $\theta$ represents the model parameters, and $N$ is the number of training examples.
During the fine-tuning process, we update the model's parameters by training it for a few epochs on the task-specific dataset. 
This enables the model to adapt to the unique characteristics and nuances of the data, resulting in improved performance on automated code update recommendation tasks.

\subsection{Recommending Code Updates}
Once the CodeLMs have been fine-tuned for the automated code recommendation dataset,  the next step is to generate code updates based on the learned representations. 
Given an input code snippet $X_i$, the fine-tuned CodeLM model generates a probability distribution over the possible updates. 
To select the most appropriate updates, the fine-tuned model employs beam search, which considers multiple code update candidates for an input sequence at each timestep based on conditional probabilities.
Once the beam search is complete, the top-ranked update candidates are selected as the final recommendations. 
These recommendations are ranked according to their probabilities, allowing developers to choose the most suitable update for their specific needs.

%% file: tables/dataset.tex
\begin{table*}[t]
  \centering
  \small
  \caption{An overview statistis of the studied datasets }
    \begin{tabular}{lrrrr}
    \toprule
    \textbf{Dataset} & \textbf{\# Train} & \textbf{\# Validation} & \textbf{\# Test} & \textbf{\# Total} \\
    \midrule
    Android &          14,690  &                1,836  &           1,835  &          18,361  \\
    Google &            9,899  &                1,237  &           1,235  &          12,371  \\
    Ovirt &          21,506  &                2,686  &           2,688  &          26,880  \\
    AndroZooUpdate-S &          19,638  &                4,910  &           6,227  &          30,194  \\
    AndroZooUpdate-L &        136,932  &              34,234  &         43,714  &        209,346  \\
    \bottomrule
    \end{tabular}%
  \label{tab:section2_data_summary}%
\end{table*}%

%% file: sections/4_experiment_setup.tex
\section{Experimental Setup}
\label{sec:experiment}
In this section, we describe the experimental setup for our study.
Our experimental setup encompasses five key aspects: studied datasets, data splitting, evaluation measures, and implementation \& hyperparameter settings.

\noindent\paragraph{\textbf{\underline{Studied Datasets}}}
As mentioned in Section~\ref{sec:approach}, we evaluate the performance of automatically recommending code updates on two datasets: \textit{AndroZooUpdate} and \textit{Pornprasit~\ea}~\cite{pornprasit2023d}.
The \textit{AndroZooUpdate} dataset consists of two versions: \textit{AndroZooUpdate-L} and \textit{AndroZooUpdate-S}. \textit{AndroZooUpdate-L} contains 209,346 pairs of updated methods extracted from 3,195 open-source Android apps, with each pair including the source code of the changed methods. However, prior work~\cite{tufano2018empirical, pornprasit2023d} has noted that large methods have a long tail distribution of sequence lengths with high variance, which may present difficulties when training an NMT model. To address this issue, we created \textit{AndroZooUpdate-S}, a smaller version of the dataset that excludes updated methods with sequences longer than 50 tokens, resulting in 30,194 pairs of smaller updated methods.
For the \textit{Pornprasit~\ea}~\cite{pornprasit2023d} dataset, we use the original subsets provided in the paper, which include code updates from three software projects: \textit{Android}, \textit{Google}, and \textit{Ovirt}.
Table~\ref{tab:section2_data_summary} provided an overview of the statistics for both datasets, offering a comprehensive summary of the data used in our study to evaluate the performance of CodeLMs in automatically recommending code updates.

\paragraph{\textbf{\underline{Data Splitting}}}
To mimic realistic recommendation scenarios, we prepare two versions of each dataset: time-wise and time-ignore. These versions differ in how the data is split into training, validation, and testing sets.
In the time-wise evaluation scenario, the dataset is first sorted in chronological order based on the commit timestamps. 
This ensures that method pairs occurring earlier in time are only present in the training dataset and not in the testing dataset. Likewise, method pairs occurring later in time are only present in the testing dataset and not in the training dataset.
This approach simulates a real-world scenario where the model is trained on historical data and evaluated on future and unseen data.
On the other hand, the time-ignore evaluation scenario does not consider the temporal information of the method pairs, which is consistent with the approach used in prior studies~\cite{tufano2018empirical, thongtanunam2022autotransform, tufano2022using}. In this scenario, the dataset is randomly split into training, validation, and testing sets without considering the temporal order of the method pairs. 
For both time-wise and time-ignore evaluations, we split the data into train/validation/test sets with a proportion of 80\%/10\%/10\%, respectively.

\paragraph{\textbf{\underline{Evaluation Measures}}}
To evaluate the accuracy, we count the number of updated methods that achieve a \textbf{perfect prediction} (PP), i.e., the generated updated method exactly matches the ground-truth (i.e., the actual updated method).
In this work, we use beam search to obtain best-$k$ candidates (i.e., \emph{$k = 1, 5, 10, 15$}). 
If one of the $k$ candidates exactly matches the ground-truth, we consider the prediction is perfect.
In this work, we also use the \textbf{BLEU} and \textbf{CodeBLEU}~\cite{ren2020codebleu} score to measure the similarity between the generated predictions and the ground-truth.
BLEU is a popular metric for evaluating the quality of the generated text, initially developed for machine translation tasks~\cite{papineni2002bleu}. 
The BLEU score measures the similarity between generated predictions and the ground truth by comparing the n-gram overlap between them. 
CodeBLEU is an extension of the traditional BLEU metric specifically designed for code generation tasks.
CodeBLEU evaluates the generated code based on four criteria: syntactic similarity, semantic similarity, token similarity, and structure similarity.
It combines these four criteria to produce a single score, which provides a more comprehensive evaluation of the generated code's quality compared to the traditional BLEU score.
In our work, we use both BLEU and CodeBLEU scores, which range from 0 to 1, to evaluate the quality of the code updates generated by our approach and the baseline approaches.
We use the BLEU-4 variant, computed by considering the 4-grams in the generated text.
BLEU-4 has been previously used in other software engineering papers (e.g., \cite{tufano2018empirical, watson2020learning}).
A higher score in both BLEU and CodeBLEU (close to 1) indicates better performance and higher similarity to the ground-truth code recommendation.


\paragraph{\textbf{\underline{Implementation \& Hyperparameter Settings}}}
We obtain the studied pre-trained models from their respective publicly available repositories and use the Hugging Face library~\cite{wolf2019huggingface} to load the model weights and generate outputs.
Given the nature of our code recommendation task, we append a transformer decoder to the encoder-based models (i.e., CodeBERT) for generating code in a regressive manner. 
In contrast, the decoder (CodeGPT) or encoder-decoder (CodeT5) models directly generate code in a regressive manner. 
The newly added fully connected layers or decoders during fine-tuning are first initialized randomly and then trained (i.e., fine-tuned).
When multiple models are provided, we select the "base" version, consistent with prior studies on CodeLMs for software engineering~\cite{zeng2022extensive, fu2022linevul,li2022automating}. 
The models are fine-tuned using an NVIDIA RTX 3090 graphics card for a total of 15 epochs. We set the learning rate to 5e-5 and the batch size to 4, as these are popular settings in code generation and understanding~\cite{zeng2022extensive}.

%% file: sections/5_result.tex
\section{Experimental Results}
\label{sec:results} 

In this section, we present the results of our empirical study with respect to our five research questions.

\subsection*{\textbf{(RQ1) \rqone}}
\paragraph{\textbf{\underline{Motivation}}}
This research question aims to investigate the performance of pre-trained CodeLMs in recommending code updates in real-world scenarios. 
Previous studies~\cite{thongtanunam2022autotransform, tufano2019learning, zeng2022extensive, li2023codeeditor} have evaluated CodeLMs in time-ignore settings, where the models are trained and tested on data without considering the chronological order of the code updates. 
However, in real-world development, code updates occur in the time-wise manner. 
Therefore, this research question aims to investigate the performance of pre-trained CodeLMs in recommending code updates in both time-ignore and time-wise scenarios. By evaluating the accuracy of CodeLMs in these two scenarios, we aim to provide insights into their effectiveness and robustness in generating suitable code recommendations.

\paragraph{\textbf{\underline{Approach}}} 
To answer this RQ, we evaluate the accuracy of CodeLMs and the baseline approaches by calculating the percentage of Perfect Predictions (\%PP), BLEU, and CodeBLEU.
We employ a beam search with a beam size of 1 to generate candidate code updates.

\input{tables/RQ1_compare_with_baseline}

\begin{figure*}[t]
    \centering
    \includegraphics[width=\textwidth]{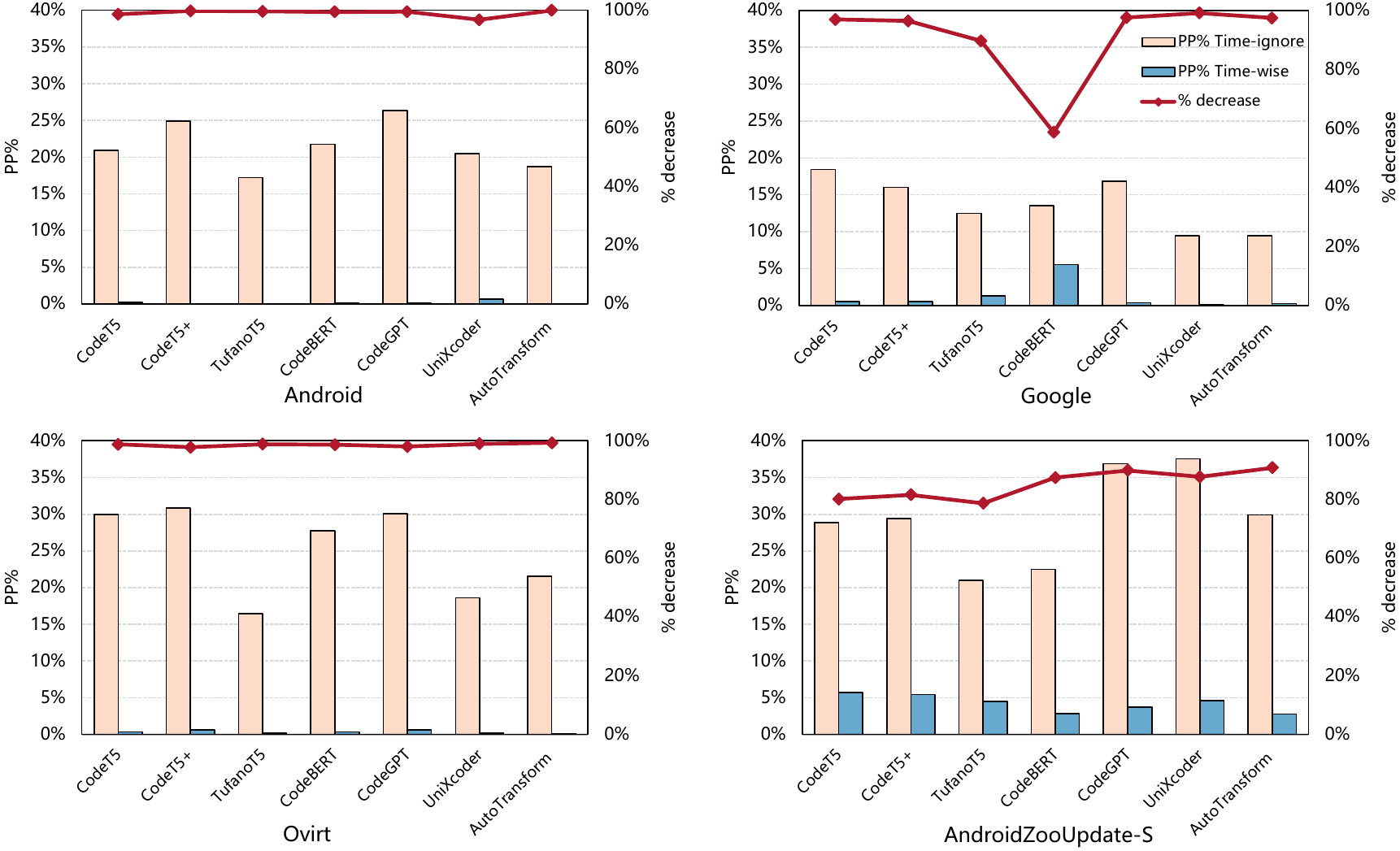}
    \caption{Comparison of CodeLMs’ perfect prediction rates in time-ignore and time-wise scenarios}
    \label{fig:rq_time_compare_senarios}
\end{figure*}

\paragraph{\textbf{\underline{Results}}} 
Table~\ref{tab:RQ1compare} presents the performance of CodeLMs on four code update datasets in time-ignore scenarios. 
First, compared to the baseline approach TufanoT5, the pre-trained CodeT5 and CodeT5+ achieve higher accuracy across all datasets. 
Although the performance varies across different datasets and models, we observe that these CodeLMs in time-ignore scenarios can achieve perfect prediction rates of almost 30\% at a beam search size of 1. 
These findings are consistent with prior research~\cite{thongtanunam2022autotransform, tufano2019learning, zeng2022extensive, li2023codeeditor} that also focused on automated code updates in time-ignore settings.

To further investigate the impact of temporal information on the performance of CodeLMs, we compare their accuracy in time-ignore and time-wise scenarios. 
Figure~\ref{fig:rq_time_compare_senarios} presents the \%PP in these two scenarios, and we calculated the decrease in \%PP using the formula:
\begin{equation*}
\text{Decrease in \%PP} = \frac{(PP_\text{time-ignore} - PP_\text{time-wise})}{PP_\text{time-ignore}} \times 100\%
\end{equation*}
The results reveal a significant decrease in performance when moving from time-ignore to time-wise settings across all CodeLMs and datasets. 
Compared to time-ignore settings, CodeLMs perform poorly in time-wise scenarios. In the Android, Google, and Ovirt datasets, the \%PP at a beam search size of 1 is close to 1\%, which is extremely low. 
The decrease rates range from 58.7\% to 100\%, highlighting the limitations of previous research that focused solely on time-ignore scenarios. 
This finding highlights the limitations of previous research that solely relied on time-ignore evaluations and emphasizes the need for more realistic assessments of CodeLMs' performance.

\begin{figure*}[t]
    \centering
    \includegraphics[width=\textwidth]{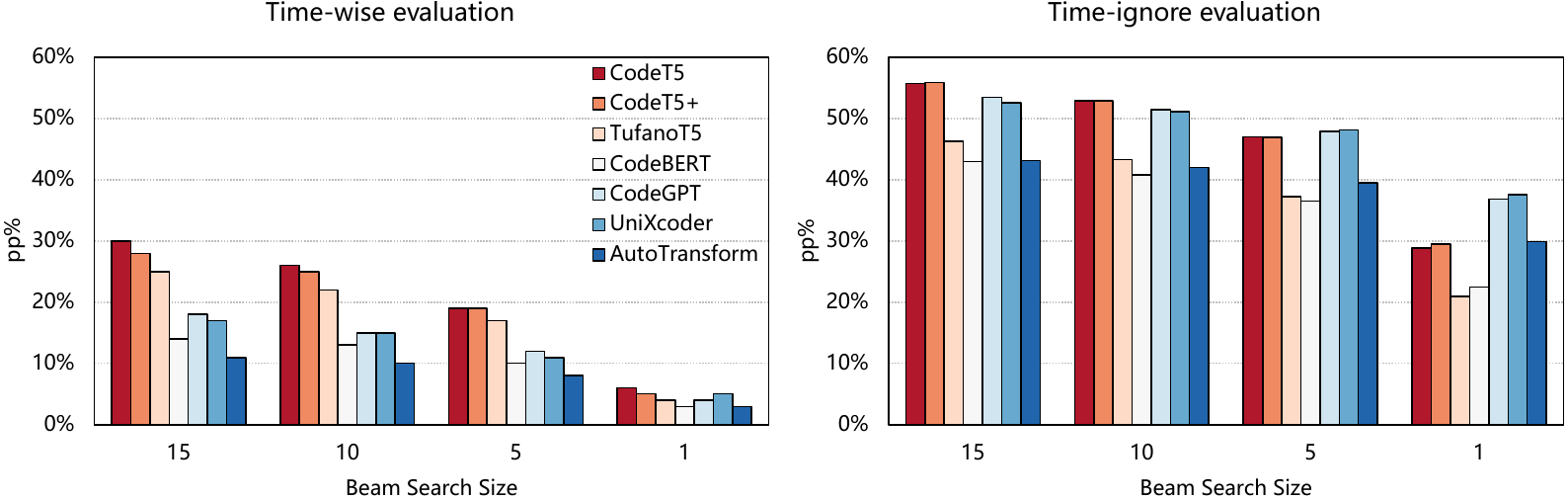}
    \caption{Impact of beam search size on CodeLMs’ performance for~\textit{AndroZooUpdate-S}}
    \label{fig:rq1_beam_search_size}
\end{figure*}

To mitigate the low accuracy in time-wise scenarios, we explore the impact of increasing the beam search size. 
Figure~\ref{fig:rq1_beam_search_size} demonstrates the effect of varying beam sizes (1, 5, 10, and 15) on the \%PP of CodeLMs in both time-ignore and time-wise scenarios for the AndroZooUpdates-S dataset. 
The results show that increasing the beam size can improve accuracy, especially in time-ignore scenarios. 
For instance, CodeT5 achieves a \%PP of 56\% with a beam size of 15 in the time-ignore scenario, compared to 29\% with a beam size of 1. 
However, the improvements are less pronounced in time-wise scenarios, with CodeT5 reaching a \%PP of 30\% with a beam size of 15, compared to 6\% with a beam size of 1. 
While increasing the beam size can improve accuracy, it may not be practical in real-world scenarios due to computational constraints and the need for developers to manually review multiple candidate updates.

\begin{figure*}[t]
    \centering
    \includegraphics[width=\textwidth]{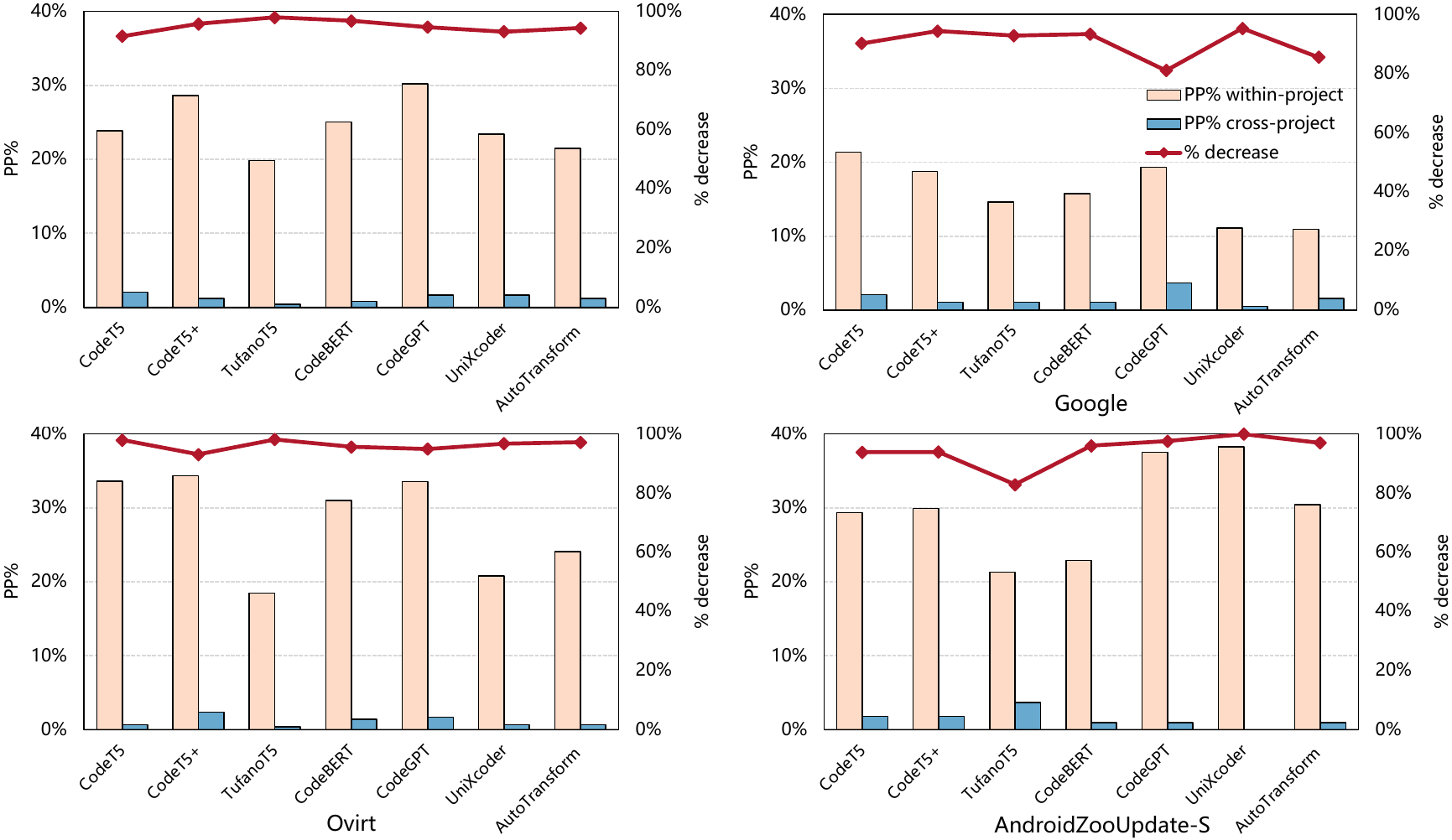}
    \caption{Comparison of CodeLMs’ PP\% for within-project and cross-project code updates in the time-ignore setting}
    \label{fig:rq_within_project}
\end{figure*}

We further investigate the generalizability of CodeLMs by comparing their performance on within-project and cross-project code updates in time-ignore settings.
Figure~\ref{fig:rq_within_project} presents the \%PP of CodeLMs for each dataset in time-ignore settings, calculated based on whether the test instances’ project is also present in the training data (within-project) or not (cross-project).
In the within-project scenario, the test instances are from projects that are also represented in the training data, while in the cross-project scenario, the test instances are from projects that are not present in the training data.
The results show a significant decrease in performance when moving from within-project to cross-project settings across all CodeLMs and datasets.
The decrease rates range from 81.0\% to 98.2\%, indicating that CodeLMs struggle to generalize to unseen projects, even in time-ignore settings.
This finding highlights the importance of considering the generalizability of CodeLMs in real-world scenarios, where code updates often involve multiple projects with different coding styles and conventions.

In summary, our results show that CodeLMs perform well in time-ignore scenarios but struggle in more realistic time-wise evaluations and cross-project code updates.
The significant decrease in accuracy when considering temporal information highlights the need for further research to improve the robustness and generalizability of CodeLMs. 
The varying performance across different datasets and models motivates our subsequent analyses to understand the factors contributing to CodeLMs’ successes and failures in recommending code updates. 
The following sections will delve deeper into these aspects, providing a more comprehensive understanding of the capabilities and limitations of CodeLMs in recommending code updates.

\find{
\textbf{Good:} CodeLMs perform well on time-ignore scenarios, achieving up to 36.9\% accuracy, and increasing the beam search can improve their performance.

\textbf{Bad:} CodeLMs perform poorly in realistic time-wise evaluations and cross-project code updates, with accuracy decreases ranging from 58.7\% to 100\%.

\textbf{Missing:} Further research is needed to understand the factors contributing to CodeLMs' low accuracy in time-wise and cross-project scenarios and to develop strategies for improving their robustness and generalizability in real-world applications.
}

\begin{figure*}[t]
    \centering
    \includegraphics[width=\textwidth]{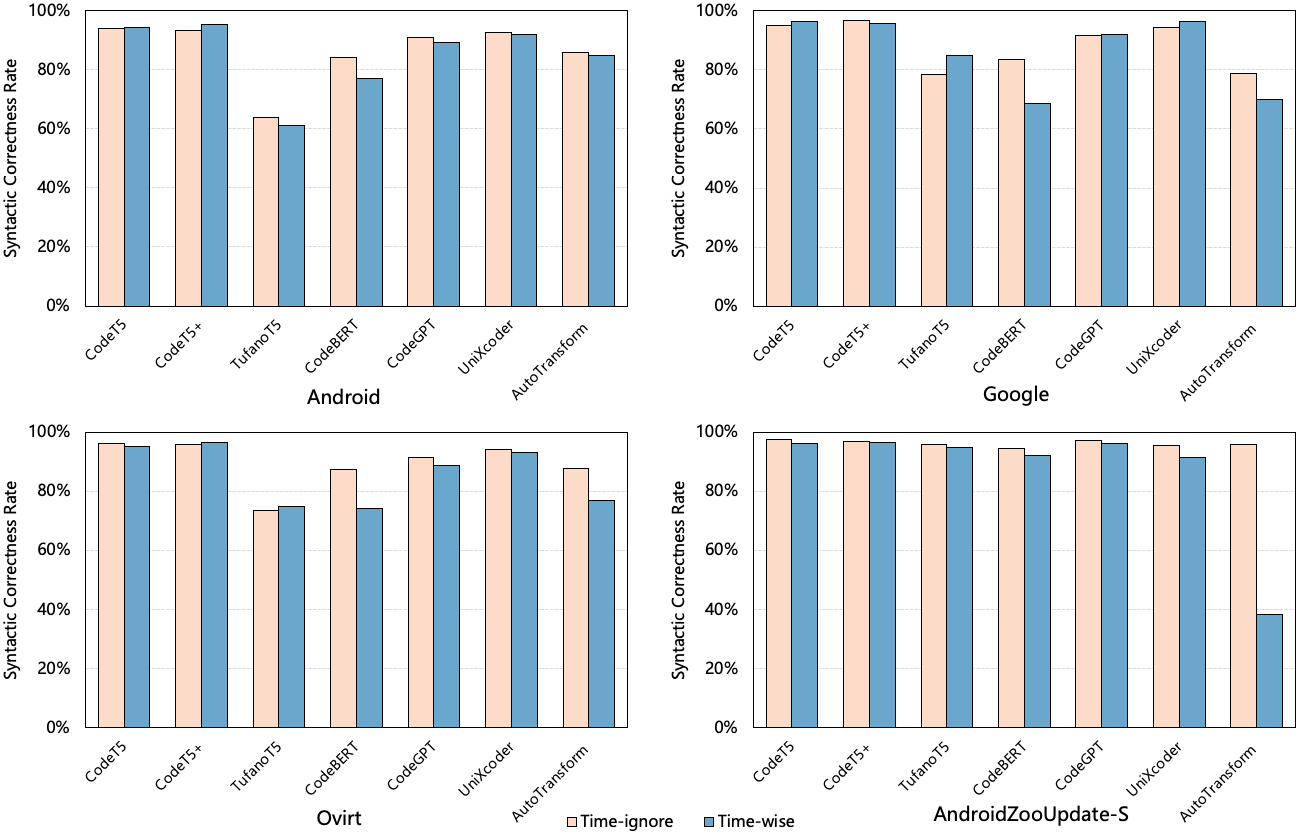}
    \caption{Syntactical correctness (in percentage) of the code generated by CodeLMs}
    \label{fig:rq2_Syntactical}
\end{figure*}

\subsection*{\textbf{(RQ2) \rqtwo}}
\paragraph{\textbf{\underline{Motivation}}}
While accuracy metrics (like perfect prediction rate) are important for evaluating CodeLMs on code updates, they don’t guarantee that the generated code is syntactically correct or functional.
Therefore, this research question aims to investigate whether the code generated by CodeLMs passes syntactic checking, providing insights into the reliability and quality of the recommended code updates.

\paragraph{\textbf{\underline{Approach}}}
To assess the syntactical correctness of the generated code, we conduct syntactic checking on the generated programs for each instance in the testing dataset. 
Following prior work~\cite{liu2023refining, kim2007prioritizing}, we utilize PMD~\cite{copeland2005pmd}, a cross-language static code analysis tool that identifies programming errors. We disregard style issues and consider an instance syntactically correct if it successfully passes the checking process.

\paragraph{\textbf{\underline{Results}}} 
Figure \ref{fig:rq2_Syntactical} presents the syntactical correctness rate of the code generated by CodeLMs across the four datasets in both time-ignore and time-wise scenarios. 
The results show that, in general, the state-of-the-art CodeLMs achieve high syntactical correctness rates across all datasets, with most models surpassing 90\% correctness in both time-ignore and time-wise scenarios.
However, it is important to note that even with high syntactical correctness rates, the generated code is not always perfect. 
There are still instances where the CodeLMs produce syntactically incorrect code. 
For example, a common issue is the mismatching of parentheses, such as generating unbalanced or excessive parentheses like "\textit{return (((((((((((((((((((}", even for better-performing models like CodeT5+.
These errors highlight the need for further improvements in the CodeLMs' ability to generate syntactically correct code.
Comparing the syntactical correctness rates between time-ignore and time-wise scenarios, we observe that the rates are generally similar for most CodeLMs. This suggests that the temporal aspect does not significantly impact the syntactical correctness of the generated code. However, there are a few exceptions, such as AutoTransform in the AndroZooUpdate-S dataset, where the syntactical correctness rate drops from 96.0\% in the time-ignore scenario to 38.2\% in the time-wise scenario. This substantial decrease may be attributed to the model’s sensitivity to temporal information or its limited ability to capture the temporal dependencies present in the AndroZooUpdate-S dataset.

\find{
\textbf{Good:} State-of-the-art CodeLMs achieve high syntactical correctness rates, often above 90\%, in most datasets and scenarios.

\textbf{Bad:} Even with high syntactical correctness rates, CodeLMs still generate syntactically incorrect code in some instances, such as mismatching parentheses.

\textbf{Missing:} Future work should focus on maintaining and improving the syntactical correctness of code generated by CodeLMs.
}

\subsection*{\textbf{(RQ3) \rqthree}}
\paragraph{\textbf{\underline{Motivation}}}
From the results of RQ1, we found that CodeLMs perform significantly differently in time-wise and time-ignore scenarios, with a substantial decrease in accuracy when considering temporal information. 
This raises the question of why there is such a big difference in performance between these two scenarios and what factors contribute to this discrepancy. 
By investigating the underlying reasons, we can gain insights into the limitations of current CodeLMs and identify potential areas for improvement in recommending code updates.

\paragraph{\textbf{\underline{Approach}}}
To understand the relationship between training and test splits, we focus on examining the similarity between the two sets at a fine-grained level. 
Although the benchmark datasets were cleaned to remove duplications, this process typically involves a high-level comparison rather than a detailed analysis of individual code pairs.
In real-world scenarios, code updates often involve a series of similar changes, such as consistently updating a variable name throughout the codebase. 
To capture these nuances and gain a deeper understanding of the similarities between training and test sets, we obtain simplified versions of each method pair.

\begin{figure*}[t]
    \centering
    \includegraphics[width=0.9\textwidth]{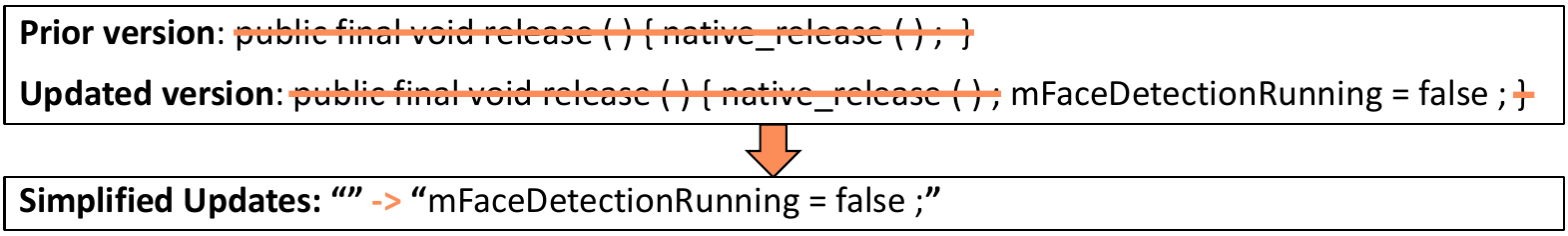}
    \caption{An Example of Update Simplification}
    \label{fig:exp_update_simplify}
\end{figure*}

As shown in Figure~\ref{fig:exp_update_simplify}, we remove the same code tokens between the prior version and the updated version code, keeping only the simplified updates for each method pair. 
This process is applied from the start and end of each method, respectively. 
For example, in Figure~\ref{fig:exp_update_simplify}, the method pair is simplified to \textit{“” -> “mFaceDetectionRunning = false;”}.
In this way, we examine the similarity between training and test sets more closely. 
By comparing the distribution of seen (simplified test samples in the training) and unseen (simplified test samples not in the training) testing samples, we aim to provide a more nuanced understanding of the factors contributing to the performance difference between time-wise and time-ignore scenarios.

\input{tables/RQ3_time_impacts}

\begin{figure*}[t]
    \centering
    \includegraphics[width=\textwidth]{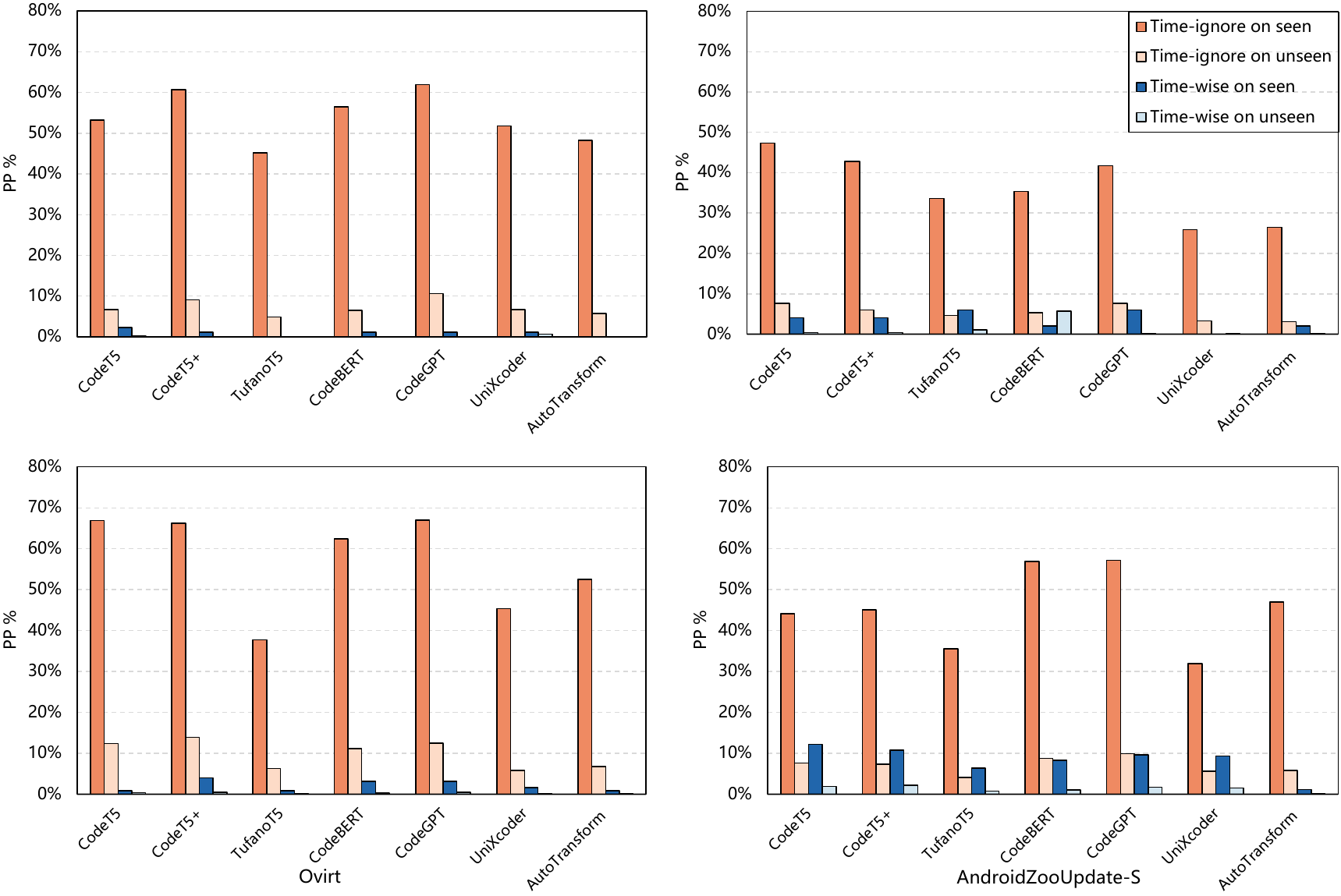}
    \caption{Perfect prediction rates of CodeLMs on seen and unseen simplified test samples}
    \label{fig:rq_seen_unseen}
\end{figure*}

\paragraph{\textbf{\underline{Results}}} 
Table~\ref{tab:RQ3_compare_time_distribution} presents the distribution of seen and unseen simplified test samples in the time-ignore and time-wise scenarios for the four datasets. 
In the time-ignore scenario, the percentage of seen examples ranges from 17.1\% to 52.4\%, indicating a significant overlap between the training and test sets. 
In contrast, the time-wise scenario exhibits a much lower percentage of seen examples, ranging from 1.6\% to 28.2\%. 
The decrease in the percentage of seen examples from time-ignore to time-wise scenarios is consistent with the decrease in performance observed in Figure~\ref{fig:rq_time_compare_senarios} in RQ1. 
Furthermore, these seen examples in the time-ignore scenario usually originate from within the same project, suggesting that CodeLMs may be relying on project-specific similar patterns rather than learning generalizable code update patterns.

Figure~\ref{fig:rq_seen_unseen} further illustrates the impact of seen and unseen simplified test samples on the \%PP of CodeLMs. In the time-ignore scenario, CodeLMs achieve significantly higher \%PP on seen examples compared to unseen ones. For instance, in the AndroidZooUpdate-S dataset, CodeT5 achieves a \%PP of 46.2\% on seen examples, while only 9.9\% on unseen examples. This trend is consistent across all datasets and CodeLMs, highlighting the reliance of CodeLMs on the presence of similar examples in the training.
In the time-wise scenario, the \%PP on both seen and unseen examples is generally much lower compared to the time-ignore scenario. However, CodeLMs still perform relatively better on seen examples than unseen ones. For example, in the Google dataset, CodeT5+ achieves a \%PP of 10.0\% on seen examples, while only 0.4\% on unseen examples.
We also manually checked the top-seen simplified updates, such as removing nonNull \textit{"@ NonNull -> "} or changing the \textit{“public -> private”}, and found that they can achieve accuracy as high as almost 90\%, as these kinds of simple changes are more common and frequent in the code base. 
These results indicate that while CodeLM-based models are capable of handling common and frequently seen code updates, they struggle with novel or complex updates that require a deeper understanding of the code’s temporal evolution.

\find{
\textbf{Good:} CodeLMs achieve high accuracy on seen test updates, especially in time-ignore scenarios.

\textbf{Bad:} CodeLMs struggle with novel or complex code updates, as shown by the low accuracy on unseen simplified test updates.

\textbf{Missing:} Future research should focus on improving CodeLMs' ability to adapt to time-wise updates and generate creative and unseen code updates.
}

\begin{figure*}[t]
    \centering
    \includegraphics[width=\textwidth]{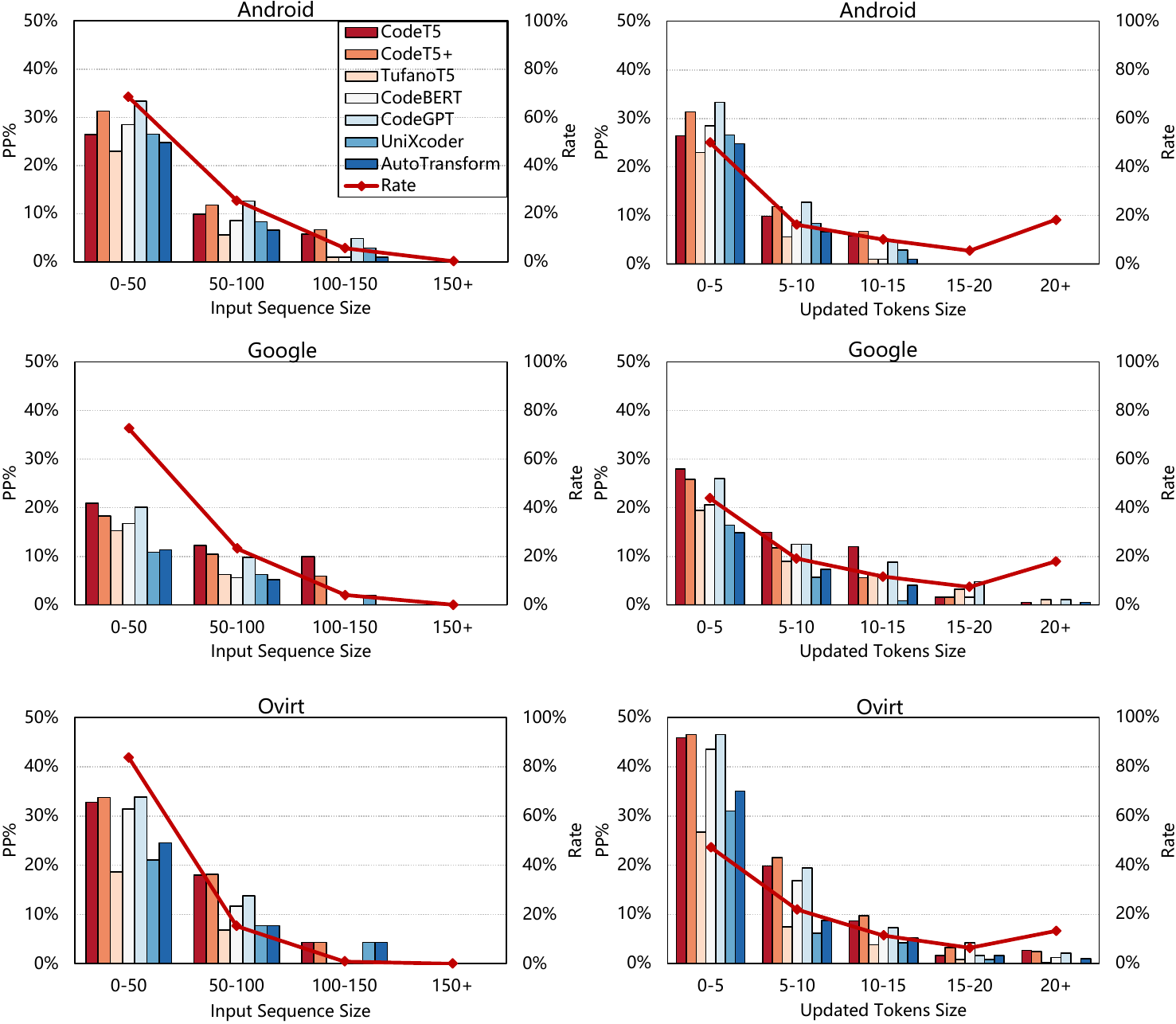}
    \caption{The impacts of  method size and the update size. }
    \label{fig:method_size_update}
\end{figure*}


\subsection*{\textbf{(RQ4) \rqfour}}

\paragraph{\textbf{\underline{Motivation}}}
In RQ1, we observed that the performance of CodeLMs was higher on AndroidZoo-Update-S, which consists of small method pairs, compared to the other three datasets.
However, little is known about the accuracy of CodeLMs on pairs of larger methods and on larger or more complex updates (i.e., updates with more changed tokens between versions of a method). 
Therefore, this research question aims to investigate the impact of method size and update size on the performance of CodeLMs in recommending code updates.

\paragraph{\textbf{\underline{Approach}}}
For the Android, Google, and Ovirt datasets, we analyze the perfect prediction rates according to the method size (i.e., the number of tokens in the initial version) and the update size (i.e., the number of changed tokens between two versions) in time-ignore scenarios. 
For AndroidZooUpdate-S, since we have limited the token size to 50, we use AndroidZooUpdate-L to assess the impact of method and update sizes on CodeLMs' performance. Due to the dataset's size and the time-consuming nature of training, we focus on the best-performing models, CodeT5 (a pre-trained CodeLM), and AutoTransform (a normal Transformer without pre-training) for this analysis. We train, validate, and test both CodeT5 and AutoTransform accordingly.
We further analyze the perfect prediction rates according to the method size and the update size at a beam size of 5.

\paragraph{\textbf{\underline{Results}}}
Figure~\ref{fig:method_size_update} presents the impact of method size and update size on the \%PP of CodeLMs for the Android, Google, and Ovirt datasets. The results demonstrate that the majority of the method pairs (69\% to 84\%) fall within the 0-50 token range, and CodeLMs achieve their highest \%PP on these smaller methods. However, as the method size increases, the proportion of method pairs decreases, and the \%PP of CodeLMs drops significantly, reaching 0\% for methods with more than 150 tokens. A similar trend is observed for update sizes, with CodeLMs performing better on updates with fewer changed tokens (0-5) and struggling with larger updates (>20 changed tokens).
This trend is consistent across all CodeLMs and datasets, highlighting the challenges faced by CodeLMs in handling larger and more complex code updates.

\input{tables/RQ4_token_size_compare}

We also evaluate the accuracy of CodeT5 and AutoTransform when considering the AndroidZooUpdate-L dataset, which includes method pairs of various sizes. CodeT5 achieves a \%PP of 8\%, while AutoTransform achieves a \%PP of 4\%, demonstrating the superiority of pre-trained CodeLMs over normal Transformers without pre-training.
Table~\ref{tab:RQ4_token_length} provides a more detailed analysis of CodeT5's accuracy according to the method size and update size on the AndroidZooUpdate-L dataset. The results show that CodeT5 performs better on smaller methods and updates, with a \%PP of 15\% for methods with 0-50 tokens and updates with 0-5 changed tokens. However, as the method size and update size increase, the \%PP decreases. For methods with more than 200 tokens and updates with more than 25 changed tokens, the \%PP drops to 1\%. Additionally, the BLEU and CodeBLEU scores also decrease as the method and update sizes increase, indicating a decline in the quality of the generated code updates.

In summary, although our previous experiments show that CodeLMs achieve much higher performance in time-ignore scenarios, their accuracy is still limited when dealing with larger methods and more complex updates. Even in time-ignore settings, the \%PP of CodeLMs approaches zero for methods with more than 150 tokens and updates with more than 20 changed tokens. This highlights the need for further research to develop CodeLMs that can effectively handle larger and more complex code updates, as these are common in real-world software development scenarios.

\find{
\textbf{Good:} CodeLMs perform better on smaller methods and updates with fewer changed tokens.

\textbf{Bad:} CodeLMs' accuracy decreases significantly as method size and update complexity increase, dropping to 0\% for larger methods and more complex updates.

\textbf{Missing:} Future research should focus on developing CodeLMs that can effectively handle larger and more complex code updates prevalent in real-world scenarios.
}

\begin{figure*}[t]
    \centering
    \includegraphics[width=\textwidth]{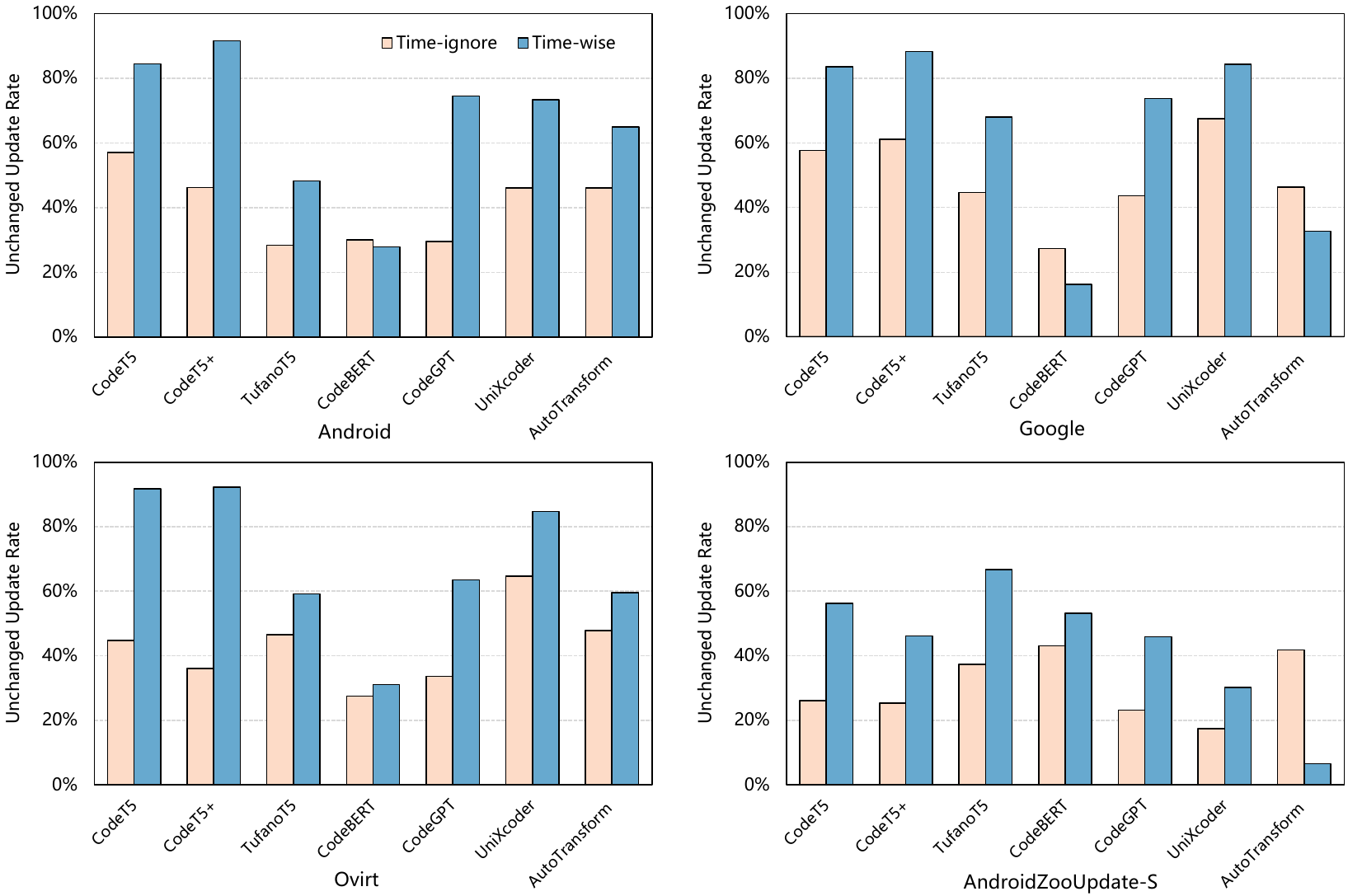}
    \caption{Rate of unchanged updates for each CodeLM across the four datasets.}
    \label{fig:rq7_unchanged_updates}
\end{figure*}

\subsection*{\textbf{(RQ5) \rqfive}}

\paragraph{\textbf{\underline{Motivation}}}
While RQ1 to RQ4 focus on the perfect prediction rates (\%PP) of CodeLMs in recommending code updates, this final research question aims to provide a qualitative analysis by examining the detailed code updates recommended by CodeLMs. By investigating the types of code updates learned and predicted by CodeLMs, we can gain insights into their strengths and limitations in capturing and generating specific code changes.

\paragraph{\textbf{\underline{Approach}}}
To analyze the types of code updates learned and predicted by CodeLMs, we employ a similar method to RQ3. 
However, instead of focusing on the simplified updates within testing and training sets, we examine the simplified updates between model inputs and model predictions. 
By removing the identical code tokens between the input and output, we obtain simplified versions of the code updates recommended by CodeLMs. This allows us to identify and categorize the types of code changes captured by the models.

\input{tables/rq7_examples_of_top_patterns}

\paragraph{\textbf{\underline{Results}}} 
By simplifying the input and output code, we found that a significant portion of the updates recommended by CodeLMs are unchanged, meaning that the models output the same content as the input. 
As shown in Figure~\ref{fig:rq7_unchanged_updates}, it is evident that the rate of unchanged updates is remarkably high, especially under the time-wise scenarios. 
In the Android dataset, CodeT5+ exhibits an unchanged update rate of 91.7\% in the time-wise scenario, while CodeT5 and UniXcoder have rates of 84.5\% and 73.4\%, respectively. 
Similarly, in the Google dataset, CodeT5+ and UniXcoder have high rates of unchanged updates (88.4\% and 84.4\%, respectively) in the time-wise scenario. 
These findings align with our previous discussions, where we observed that the accuracy of CodeLMs is much lower under the time-wise scenarios compared to the time-ignore scenarios.
Interestingly, even in the time-ignore scenarios, the rate of unchanged updates is still substantial. 
For instance, in the Google dataset, CodeT5 and CodeT5+ have unchanged updates almost 60\%.
These findings suggests that face challenges in recommending meaningful code changes.

To further investigate the types of code updates learned and predicted by CodeLMs, we examine the top simplified code updates in each dataset. Table~\ref{tab:rq7_android_update_examples} presents the top 10 simplified code updates recommended by CodeLMs for the Android dataset under the time-ignore scenario, along with the percentage of the update existing in the training set and the relevant \%PP. 
For the Android dataset, each model generates 1,836 updates.
From Table~\ref{tab:rq7_android_update_examples}, we observe that CodeLMs can accurately predict certain types of code updates, such as converting "os" to "system" and "libcore.io" to "android.system". 
These updates are relatively simple and occur frequently in the training set, enabling the models to learn and predict them effectively. 
For example, the update \textit{"os" -> "system"} exists in 2.2\% of the training set, and we also checked that these test examples are from the same repository. Code~\ref{lst:v2code2} shows an example of this type of update, where the package name is changed from "android.os" to "android.system".

\begin{lstlisting}[language=diff, label=lst:v2code2, caption = Examples of code update (“os” -> “system”). ]
- public android.os.StructUtsname uname() {
+ public android.system.StructUtsname uname() {
- public void setegid(int egid) throws android.os.ErrnoException;
+ public void setegid(int egid) throws android.system.ErrnoException;
\end{lstlisting}

\find{
\textbf{Good:} CodeLMs accurately predict simple and frequent code updates.

\textbf{Bad:} CodeLMs struggle to recommend meaningful changes, especially in time-wise scenarios, resulting in high rates of unchanged updates. 

\textbf{Missing:} Future research should focus on improving CodeLMs' adaptability and generalization capabilities to recommend meaningful changes in the presence of new and unseen code patterns.
}

%% file: tables/RQ1_compare_with_baseline.tex
\begin{table}[t]
  \centering
  \caption{Performance of CodeLMs on the time-ignore setting}
  \resizebox{\linewidth}{!}{
    \begin{tabular}{lcccccccccccc}
    \toprule
    \multirow{2}[2]{*}{\textbf{Approach}} & \multicolumn{3}{c}{\textbf{Android}} & \multicolumn{3}{c}{\textbf{Google}} & \multicolumn{3}{c}{\textbf{Ovirt}} & \multicolumn{3}{c}{\textbf{AndroZooUpdate-S}} \\
          & \textbf{PP\%} & \textbf{BLEU} & \textbf{CodeBLEU} & \textbf{PP\%} & \textbf{BLEU} & \textbf{CodeBLEU} & \textbf{PP\%} & \textbf{BLEU} & \textbf{CodeBLEU} & \textbf{PP\%} & \textbf{BLEU} & \textbf{CodeBLEU} \\
    \midrule
    \textbf{CodeT5} & 20.9\% & 0.86  & 0.85  & 18.4\% & 0.86  & 0.86  & 30.0\% & 0.86  & 0.87  & 28.9\% & 0.95  & 0.93 \\
    \textbf{CodeT5+} & 24.9\% & 0.87  & 0.86  & 16.0\% & 0.86  & 0.86  & 30.8\% & 0.85  & 0.87  & 29.4\% & 0.93  & 0.92 \\
    \textbf{TufanoT5} & 17.2\% & 0.81  & 0.84  & 12.5\% & 0.78  & 0.83  & 16.5\% & 0.67  & 0.81  & 21.0\% & 0.76  & 0.84 \\
    \textbf{CodeBERT} & 21.8\% & 0.86  & 0.85  & 13.5\% & 0.81  & 0.82  & 27.8\% & 0.84  & 0.85  & 22.5\% & 0.95  & 0.92 \\
    \textbf{CodeGPT} & 26.4\% & 0.85  & 0.85  & 16.9\% & 0.83  & 0.83  & 30.1\% & 0.84  & 0.85  & 36.9\% & 0.96  & 0.94 \\
    \textbf{UniXcoder} & 20.5\% & 0.86  & 0.85  & 9.5\% & 0.85  & 0.85  & 18.6\% & 0.83  & 0.86  & 37.6\% & 0.94  & 0.93 \\
    \textbf{AutoTransform } & 18.7\% & 0.87  & 0.86  & 9.5\% & 0.85  & 0.85  & 21.5\% & 0.85  & 0.86  & 29.9\% & 0.96  & 0.93 \\
    \bottomrule
    \end{tabular}}%
  \label{tab:RQ1compare}%
\end{table}%

%% file: tables/RQ3_time_impacts.tex
\begin{table*}[t]
  \centering
  \caption{Distribution of seen and unseen simplified test samples in the datasets}
    \small
    \begin{tabular}{lrrrrc}
    \toprule
          & \multicolumn{2}{c}{\textbf{Time-ignore}} & \multicolumn{2}{c}{\textbf{Time-wise}} &  \\
          & \textbf{Seen} & \textbf{Unseen} & \textbf{Seen} & \textbf{Unseen} & \multicolumn{1}{r}{\textbf{\% decrease (seen)}} \\
    \midrule
    \textbf{Android} & 30.6\% & 69.4\% & 4.7\% & 95.3\% & 84.5\% \\
    \textbf{Google} & 27.2\% & 72.8\% & 4.0\% & 96.0\% & 85.1\% \\
    \textbf{Ovirt} & 32.4\% & 67.6\% & 4.8\% & 95.2\% & 85.3\% \\
    \textbf{AndroidZooUpdate-S} & 58.5\% & 41.5\% & 37.5\% & 62.5\% & 35.9\% \\
    \bottomrule
    \end{tabular}%
  \label{tab:RQ3_compare_time_distribution}%
\end{table*}%

%% file: tables/RQ4_token_size_compare.tex
{\Large
\begin{table*}[t]
  \centering
  \large
  \caption{The accuracy of CodeT5 according to the method size and the update size.}
  \resizebox{\linewidth}{!}{

    \begin{tabular}{cccccccccccccccc}
    \toprule
    \textbf{Method size} & \multicolumn{3}{c}{\textbf{0-50}} & \multicolumn{3}{c}{\textbf{50-100}} & \multicolumn{3}{c}{\textbf{100-150}} & \multicolumn{3}{c}{\textbf{150-200}} & \multicolumn{3}{c}{\textbf{200+}} \\
    \midrule
    \textbf{\#Updated tokens} & \textbf{PP\%} & \textbf{\B} & \textbf{\CB} & \textbf{PP\%} & \textbf{\B} & \textbf{\CB} & \textbf{PP\%} & \textbf{\B} & \textbf{\CB} & \textbf{PP\%} & \textbf{\B} & \textbf{\CB} & \textbf{PP\%} & \textbf{\B} & \textbf{\CB} \\
    \midrule
    \textbf{0-5} & 15\%  & 0.39  & 0.75  & 12\%  & 0.88  & 0.89  & 11\%  & 0.94  & 0.92  & 11\%  & 0.96  & 0.93  & 5\%   & 0.35  & 0.48 \\
    \textbf{5-10} & 8\%   & 0.42  & 0.70  & 8\%   & 0.87  & 0.85  & 8\%   & 0.93  & 0.89  & 8\%   & 0.95  & 0.91  & 5\%   & 0.52  & 0.60 \\
    \textbf{10-15} & 9\%   & 0.38  & 0.66  & 7\%   & 0.82  & 0.81  & 7\%   & 0.92  & 0.87  & 5\%   & 0.94  & 0.89  & 3\%   & 0.52  & 0.60 \\
    \textbf{15-20} & 6\%   & 0.39  & 0.65  & 9\%   & 0.81  & 0.79  & 8\%   & 0.89  & 0.84  & 7\%   & 0.92  & 0.87  & 3\%   & 0.44  & 0.54 \\
    \textbf{20-25} & 7\%   & 0.38  & 0.61  & 4\%   & 0.78  & 0.74  & 6\%   & 0.87  & 0.81  & 4\%   & 0.91  & 0.85  & 1\%   & 0.48  & 0.56 \\
    \textbf{25+} & 5\%   & 0.24  & 0.62  & 2\%   & 0.54  & 0.61  & 3\%   & 0.70  & 0.66  & 2\%   & 0.71  & 0.68  & 1\%   & 0.02  & 0.24 \\
    \bottomrule
    \end{tabular}
}
\begin{flushleft}
  \footnotesize
  \textit{Note: B represents BLEU score and CB represents CodeBLEU score.}
  \end{flushleft}
  \label{tab:RQ4_token_length}%
\end{table*}%
}

%% file: tables/rq7_examples_of_top_patterns.tex
\begin{table}[t]
  \centering
  \caption{Top 10 Simplified Code Updates Recommended by CodeLMs in Android Dataset}
  \resizebox{\linewidth}{!}{
    \begin{tabular}{lrrrrrrrrr}
    \toprule
    \textbf{Simplified Code Update} & \multicolumn{1}{l}{\textbf{Training\%}} & \textbf{CodeT5} & \textbf{CodeT5+} & \textbf{TufanoT5} & \textbf{CodeBERT} & \textbf{CodeGPT} & \textbf{UniXcoder} & \textbf{AutoTransform} \\
    \midrule
    \textit{“os” -> “system”} & 2.2\% & 100\%(28) & 97\%(30) & 100\%(27) & 100\%(28) & 97\%(29) & 97\%(29) & 100\%(27) \\
    \textit{“libcore.io” -> “android.system”} & 1.3\% & 94\%(16) & 94\%(16) & 94\%(16) & 94\%(16) & 94\%(16) & 94\%(16) & 94\%(16) \\
    \textit{“Constants.RIL” -> “”} & 0.7\% & 100\%(14) & 100\%(14) & 100\%(10) & 100\%(16) & 100\%(16) & 100\%(15) & 100\%(13) \\
    \textit{“” -> “@com.android.annotations.NonNull”} & 0.4\% & 42\%(12) & 42\%(12) & 43\%(7) & 67\%(12) & 60\%(5) & 60\%(10) & 50\%(4) \\
    \textit{“” -> “noinline”} & 0.4\% & 100\%(11) & 85\%(13) & 0\%(0) & 91\%(11) & 92\%(12) & 91\%(11) & 100\%(9) \\
    \textit{“” -> “@java.lang.Override”} & 0.7\% & 64\%(11) & 100\%(6) & 100\%(6) & 82\%(11) & 100\%(7) & 67\%(6) & 69\%(13) \\
    \begin{tabular}[c]{@{}l@{}}\textit{“addProperty(Options.LAMBDA\_MODE.getName(),} \\ \textit{Options.LambdaMode.LEGACY.toString()).” -> “”}\end{tabular} & 0.6\% & 90\%(10) & 100\%(9) & 100\%(9) & 100\%(9) & 100\%(9) & 100\%(7) & 100\%(8) \\ 

    \textit{“this.” -> “”} & 0.2\% & 63\%(8) & 67\%(6) & 86\%(7) & 75\%(8) & 57\%(7) & 60\%(10) & 80\%(5) \\
    \textit{“/jack” -> “”} & 0.6\% & 100\%(7) & 100\%(7) & 100\%(7) & 100\%(7) & 100\%(7) & 100\%(7) & 100\%(7) \\
    \textit{“getKey(256)” -> “CRYPT\_KEY”} & 0.1\% & 100\%(7) & 100\%(7) & 100\%(7) & 100\%(2) & 100\%(7) & 100\%(7) & 100\%(7) \\
    \bottomrule
    \end{tabular}%
    }
  \label{tab:rq7_android_update_examples}%
\end{table}%

%% file: sections/discussion.tex
\section{Discussion}
\label{sec:discussion}

\smallsection{Efficiency}
One of the important factors when considering the adoption of CodeLMs in practice is the computation cost. 
Hence, we measure the fine-tuning and inference time for the CodeLMs.
As we described, the training and inference time is based on a computer with an NVIDIA RTX 3090 graphics card. 
We find that the fine-tuning time of CodeLMs ranges from 7 hours to 13 hours depending on the model types and the number of epochs. 
Specifically, models such as T5 and CodeT5 required an average of seven hours for fine-tuning. In contrast, models like CodeBERT and CodeGPT exhibited longer fine-tuning times, often exceeding 10 hours.
The average inference time for generating one sequence candidate per method is 0.3 seconds, which is relatively fast and acceptable.
However, when we train the CodeLMs on the larger \textit{AndroidZooUpdate-L} dataset for the same 15 epochs (RQ4), the fine-tuning time increases significantly, requiring almost four days. 
This indicates that the computation cost of CodeLMs increases dramatically with increasing method size and dataset size.
This poses a challenge to the scalability and applicability of CodeLMs for code update recommendation.
To address this challenge, future research should explore techniques for improving their efficiency, such as distillation, pruning, and quantization~\cite{hou2023large}.

\smallsection{Evaluation metrics}
In our study, we used three evaluation metrics to assess the performances of CodeLMs: \%PP, BLEU, and CodeBLEU. As shown in Table~\ref{tab:RQ1compare}, these metrics generally exhibit a consistent trend when comparing different CodeLMs under the same beam size. However, an interesting observation is that while CodeT5 only achieves a 20.9\% perfect prediction rate on Android dataset, it scores high on BLEU (0.86) and CodeBLEU (0.85). This suggests that a non-perfect prediction may still achieve a high BLEU score if it shares many tokens with the ground truth.
Although we primarily focus on the \%PP metric in our analysis, it is important to note that other aspects, such as syntactical correctness and functional equivalence, are also crucial for evaluating the quality of code update recommendations. 
To address this limitation, we suggest that future research should explore more comprehensive evaluation metrics that consider multiple dimensions of code quality. For instance, incorporating static code analysis tools to assess the syntactical correctness of the generated code updates can provide a more robust measure of their validity. Additionally, manual analysis by domain experts can offer valuable insights into the functional equivalence and semantic correctness of the recommended code updates.

%% file: sections/threats.tex
\section{Threats to Validity}
\label{sec:threats}

In this section, we discuss potential threats that may influence the outcomes of our study.

\textbf{Threats to the external validity}
concern the generalizability of our work.
Our study is based on the \textit{AndroidZooUpdate} and Pornprasit~\ea~\cite{pornprasit2023d}.
For \textit{AndroidZooUpdate}, we mainly collect Java method blocks from these Android apps.
There is a possibility that these apps may not be representative of open-source Android apps developed in the Kotlin programming language, which Google officially supported for Android development starting in 2017~\cite{KotlinWeb}.
However, Java is a widely used programming language, extensively employed in Android development~\cite{liu2020androzooopen, gois2019empirical}. 
Additionally, our selection criteria, which required apps to be published on Google Play and have more than one commit message on GitHub, led to the exclusion of a large number of apps from the AndroZooOpen dataset. While this was necessary to ensure that we were working with apps that are actively maintained, it does introduce a potential threat to the generalizability of our results. However, to mitigate this threat and maintain the generality of our findings, we have included popular benchmark datasets Pornprasit~\ea~\cite{pornprasit2023d} (i.e., Android, Google, Ovirt) in our analysis alongside our collected dataset.

\textbf{Threats to the internal validity}
relate to the impact of the hyperparameter settings on the performance of CodeLM-based models.  
During fine-tuning, we also did not alter the model's architecture (e.g., number of layers).
We acknowledge that a comprehensive calibration would certainly yield better results.
However, optimizing the settings of the parameter for deep learning approaches is computationally expensive. 
Furthermore, our research objective does not involve determining the optimal settings but rather aims to fairly investigate the feasibility of learning and recommending code updates using CodeLMs.

\textbf{Threats to the construct validity} relate to the data collection process for \textit{AndroidZooUpdate}, as it may have an impact on the accuracy of the CodeLMs.
We perform data cleaning and filtering steps to mitigate the noise in \textit{AndroidZooUpdate} (e.g., removing duplicated methods pairs).
However, unknown noise may still exist in our curated dataset. 
Future work should further investigate unknown noise and explore its impact on the evaluation of code update recommendation approaches using CodeLMs.

%% file: sections/6_relatedwork.tex
\section{Related Work}
\label{sec:relatedwork}
\subsection{Deep Learning for Software Engineering}
In the past decade, Deep Learning (DL) has emerged as a dominant research domain, bringing several breakthroughs in image, text, video, and speech processing~\cite{lecun2015deep}.
DL has also been extensively applied to natural language processing and source code analysis in software engineering~\cite{yang2020survey}.
Unlike traditional machine learning techniques that rely heavily on manually-crafted feature representations, deep learning enables the automatic learning of feature representations from raw data~\cite{lecun2015deep, goodfellow2016deep}.
Motivated by the remarkable success of deep learning techniques, researchers have demonstrated considerable interest in applying these techniques to various software engineering tasks (e.g., code search~\cite{gu2018deep}, code generation~\cite{kim2021code}, defect prediction \& localization~\cite{wattanakriengkrai2020predicting, dam2019lessons, pornprasit2021jitline, pornprasit2022deeplinedp, huo2019deep}, vulnerability prediction~\cite{fu2022vulrepair, dam2018automatic, fu2023chatgpt,  fu2024aibughunter, fu2022linevul, fu2023vulexplainer,  fu2024vision}, test case generation~\cite{alagarsamy2024enhancing,alagarsamy10a3test, watson2020learning}, code review~\cite{tufan2021towards, tufano2019learning, thongtanunam2022autotransform, pornprasit2024gpt, hong2022commentfinder, hong2022should}, CI/CD~\cite{hong2024practitioners}, and DevSecOps~\cite{fu2024ai}.
Specific to the mobile-software engineering context, researchers also proposed various automated approaches for UI generation~\cite{chen2018ui, chen2020unblind}, mobile malware classification~\cite{yuan2016droiddetector, zhang2020enhancing} and mobile privacy policy analysis~\cite{harkous2018polisis}.

\subsection{AI/ML-based Experimental Bias in Software Engineering}
According to the survey results presented by Yang~\ea~\cite{yang2020survey}, machine learning or deep learning technologies have been increasingly used in software engineering to improve developers' productivity and software quality.
However, careful consideration must be given when evaluating AI/ML-based approaches in software engineering, such as defect prediction~\cite{tantithamthavorn2018experience}, malware classification~\cite{pendlebury2019tesseract}, and large language models~\cite{she2023pitfalls}.
Prior studies have raised various potential issues that may impact the accuracy and interpretation of AI/ML models in SE.
These issues include the quality of datasets~\cite{ghotra2015revisiting,tantithamthavorn2015impact}, data labeling techniques~\cite{yatish2019mining}, feature selection techniques~\cite{ghotra2017large, jiarpakdee2018autospearman}, collinearity analysis~\cite{jiarpakdee2018autospearman, jiarpakdee2018impact, jiarpakdee2016study}, class rebalancing techniques~\cite{tantithamthavorn2018impact}, model construction~\cite{ghotra2015revisiting}, parameter optimization~\cite{agrawal2018better, agrawal2020better, fu2016tuning, tantithamthavorn2016automated, tantithamthavorn2018impact}, model evaluation~\cite{tantithamthavorn2016empirical, pornprasit2021jitline, jimenez2019importance}, and model interpretation~\cite{jiarpakdee2018impact,jiarpakdee2020empirical}.

Extending the concerns related to evaluation techniques~\cite{tantithamthavorn2016empirical, pornprasit2021jitline, jimenez2019importance}, this paper is the first to demonstrate the impact of the time-wise evaluation scenario for various CodeLM-based code generation tasks (including our own code update recommendation tool and the automated code transformation tool in prior work~\cite{tufano2019learning, thongtanunam2022autotransform}).
Thus, our finding highlights the importance of the time-wise evaluation scenario, which should be considered in future work.

%% file: sections/7_conclusion.tex
\section{Conclusion}
\label{sec:conclusion}
In this work, we investigated the application of CodeLMs, specifically CodeT5, CodeT5+, CodeBERT, CodeGPT, and UniXcoder, to the task of automatically recommending code updates.
Through an extensive evaluation on two diverse datasets, we assessed the performance of various CodeLMs across different scenarios, considering factors such as temporal evolution, project specificity, method size, and update complexity.
Our findings reveal that while CodeLMs demonstrate promising results in settings that ignore temporal information, their performance significantly declines in more realistic time-wise scenarios. 
The accuracy of CodeLMs drops by 58.7\% to 100\% when faced with the challenges of adapting to new projects and handling the temporal nature of code updates.
Furthermore, we find that CodeLMs often generate syntactically incorrect or null updates, especially for larger methods and more complex edits, highlighting their limitations in producing reliable and meaningful code changes.
Our analysis also uncovers that CodeLMs heavily rely on previously seen samples, struggling to generalize to novel and challenging updates.
These findings underscore the significant gap between the perceived and actual effectiveness of CodeLMs for practical code update recommendation. While CodeLMs have demonstrated impressive results on various software engineering benchmarks, our study reveals that their performance does not directly translate to real-world scenarios, emphasizing the need for further research and development efforts to bridge this gap and realize the full potential of CodeLMs in automating code updates.